\documentclass[conference]{IEEEtran}

\usepackage{cite}
\usepackage{amsmath,amssymb,amsfonts}
\usepackage{algorithmic}
\usepackage{graphicx}
\usepackage{textcomp}
\usepackage{xcolor}
\usepackage{multirow}
\usepackage[normalem]{ulem}
\usepackage{epsfig, subfigure, amsmath, amssymb}

\usepackage{amsmath}
\usepackage{graphicx}

\makeatletter
\def\ps@IEEEtitlepagestyle{%
  \def\@oddfoot{\mycopyrightnotice}%
  \def\@evenfoot{}%
}
\def\mycopyrightnotice{%
  {\footnotesize \makebox[\columnwidth]{978-1-7281-8192-9/21/\$31.00~\copyright2021 IEEE \hfill} \hspace{\columnsep}\makebox[\columnwidth]{ }\hfill}
  \gdef\mycopyrightnotice{}
}

\definecolor{red}{rgb}{1,0,0}

\def\BibTeX{{\rm B\kern-.05em{\sc i\kern-.025em b}\kern-.08em
    T\kern-.1667em\lower.7ex\hbox{E}\kern-.125emX}}

\title{Machine Learning in Generation, Detection, and Mitigation of Cyberattacks in Smart Grid: A Survey}

\author {\IEEEauthorblockN{Nur Imtiazul Haque, Md Hasan Shahriar, Md Golam Dastgir, Anjan Debnath, \\ Imtiaz Parvez, Arif Sarwat, Mohammad Ashiqur Rahman}
Department of Electrical and Computer Engineering\\
Florida International University, Miami, USA\\
\{nhaqu004, mshah068, mdast001, adebn001, iparv001, asarwat, marahman\}@fiu.edu}
\begin{document}

\maketitle

\begin{abstract}
Smart grid (SG) is a complex cyber-physical system that utilizes modern cyber and physical equipment to run at an optimal operating point. Cyberattacks are the principal threats confronting the usage and advancement of the state-of-the-art systems. The advancement of SG has added a wide range of technologies, equipment, and tools to make the system more reliable, efficient, and cost-effective. Despite attaining these goals, the threat space for the adversarial attacks has also been expanded because of the extensive implementation of the cyber networks. Due to the promising computational and reasoning capability, machine learning (ML) is being used to exploit and defend the cyberattacks in SG by the attackers and system operators, respectively. In this paper, we perform a comprehensive summary of cyberattacks generation, detection, and mitigation schemes by reviewing state-of-the-art research in the SG domain. Additionally, we have summarized the current research in a structured way using tabular format. We also present the shortcomings of the existing works and possible future research direction based on our investigation.

\end{abstract}


\section{Introduction}
\label{Sec:Introduction}
\begin{figure}[h]
\centering
\vspace{-5pt}
    \includegraphics[scale=0.7]{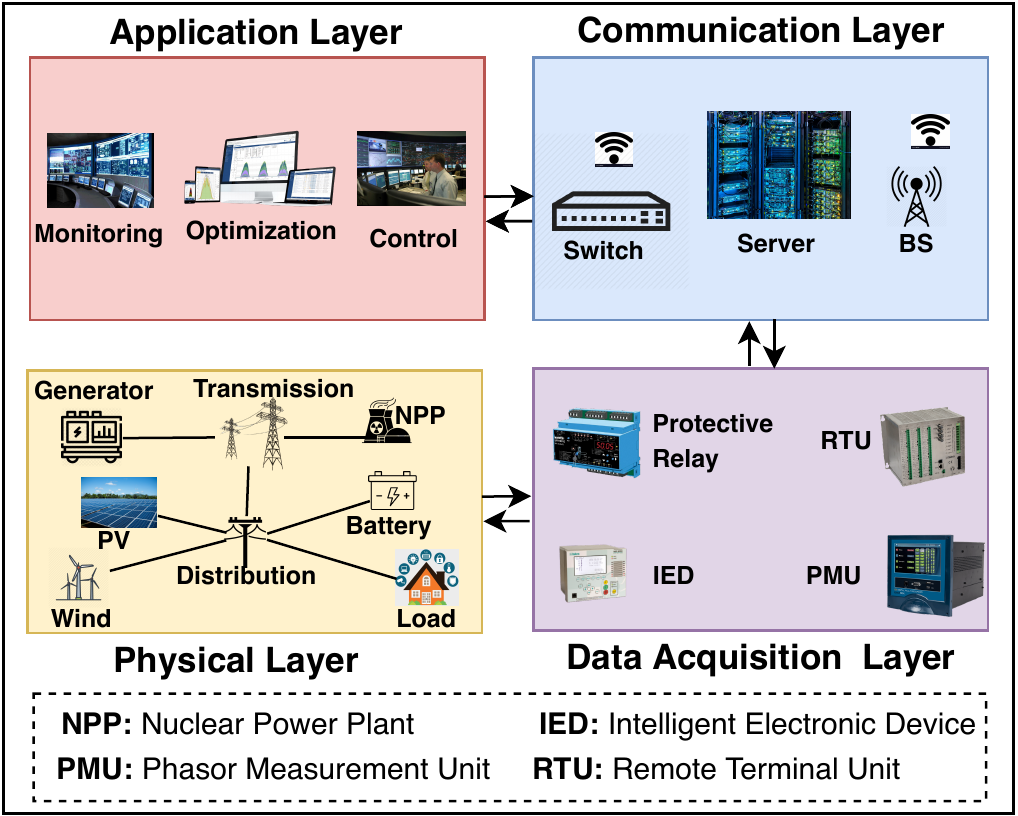}
  \vspace{-5pt}
    \caption[width = 0cm]{Architecture of the cyber-physical systems of smart grid 
    }
    \label{architecture}
\vspace{-5pt}
\end{figure}

In the modern power system, a vast amount of intelligent devices form a cyber network to monitor, control, and protect the physical network. The cyber-physical (CP) networks' inter-dependency is the backbone of the modern smart grid (SG). Fig.~\ref{architecture} illustrates the typical multi-layer architecture of SG, composed of physical, data acquisition, communication, and application layers. The physical layer consists of generation, transmission, and distribution networks~\cite{debnath2020novel,saha2020photovoltaic, jafari2019study}. SG incorporates distributed energy resources (DERs) such as solar, wind, hydro, etc., connected to the grid with converters for extracting maximum power~\cite{debnath2012fast, 7372996, debnath2020, shahriar2016stability}. The data acquisition layer consists of smart sensors and measurement devices, where the smart devices collect data and transmit them to the communication layer. The communication layer includes a wide variety of wired/wireless technologies and network devices, which transmits data to the energy management system (EMS) that optimizes, monitors, and sends control signals to the actuators.

Though the cyber layers improve the efficiency of the SG, they might put the system at higher risk by expanding the attack space. An attacker can compromise the vulnerable points, disrupting the monitoring and controlling of the physical equipment~\cite{sanjab2016smart, newaz2019healthguard, newaz2020heka, rahman2019novel}. 
Additionally, demand response, energy efficiency, dynamic electricity market, distributed automation, etc. are the key features of the SG network~\cite{faheem2018smart,parvezmulti, rahman2019false}.
All these salient features nominate the SG, a very complex network. machine learning (ML) is a ubiquitous prominent tool with capability of extracting patterns in any complex network data without being explicitly programmed. Recently, a lot of researchers are using ML to analyze the cybersecurity of SG.


\begin{table*}[!ht]
\scriptsize
\centering
\caption{Classification of ML-based attack generation techniques in smart grid\label{tab:attack}}
\begin{tabular}{|l|l|l|l|l|l|l|l|l|}
\hline
\textbf{Reference} & \textbf{Institution}                                                                                                    & \textbf{Year}          & \textbf{Attack Type}                                                        & \textbf{Attack Goal}                                                                                                  & \textbf{Category}                                    & \textbf{\begin{tabular}[c]{@{}l@{}}ML \\ Algorithm\end{tabular}} & \textbf{Performance}                                                                                               & \textbf{Testbed}                                                                           \\ \hline
Paul et al.~\cite{paul2019comparative}       & \begin{tabular}[c]{@{}l@{}}South Dakota State \\ University, USA\end{tabular}                      &                        &                                                                             & \begin{tabular}[c]{@{}l@{}}Generation loss, \\ Line outage\end{tabular}                          & Unsupervised                    & K-means                                     & \begin{tabular}[c]{@{}l@{}}Line outage-63 \\ Generation loss- \\ 12029 MW\end{tabular}       & \begin{tabular}[c]{@{}l@{}}W\&W 6, IEEE 7, \\ 8, 300 bus\end{tabular} \\ \cline{1-2} \cline{5-9} 
Ni et al.~\cite{ni2019multistage}          & \begin{tabular}[c]{@{}l@{}}South Dakota State \\ University, USA\end{tabular} & \multirow{-2}{*}{2019} &                                                                             & \begin{tabular}[c]{@{}l@{}}Optimal Multistage \\ Attack Sequence \\ for Line outage\end{tabular} &                                 &                                             & \begin{tabular}[c]{@{}l@{}}Line outage- \\ 30\%, Generation \\ loss-60\%\end{tabular}        & \begin{tabular}[c]{@{}l@{}}W\&W 6 bus and\\  IEEE 39 bus\end{tabular} \\ \cline{1-3} \cline{5-5} \cline{8-9} 
Ni et al.~\cite{ni2017reinforcement}          & \begin{tabular}[c]{@{}l@{}}South Dakota State \\ University, USA\end{tabular}                      & 2017                   &                                                                             & \begin{tabular}[c]{@{}l@{}}Optimal Attack \\ Sequence for \\ Blackout\end{tabular}                  &                                 &                                             &                                                                                               & \begin{tabular}[c]{@{}l@{}}W\&W 6 bus and \\ IEEE 30 bus\end{tabular} \\ \cline{1-3} \cline{5-5} \cline{9-9} 
Yan et al.~\cite{yan2016q}         & \begin{tabular}[c]{@{}l@{}}University of Rhode \\ Island, USA\end{tabular}                         & 2016                   & \multirow{-4}{*}{\begin{tabular}[c]{@{}l@{}}Line \\ Switching\end{tabular}} & \begin{tabular}[c]{@{}l@{}}Optimal Attack \\ Sequence for \\ Blackout\end{tabular}                  & \multirow{-3}{*}{Reinforcement} & \multirow{-3}{*}{Q-learning}                & \multirow{-3}{*}{\begin{tabular}[c]{@{}l@{}} Generation \\loss- 160.1278\\MW and blackout\end{tabular}} & \begin{tabular}[c]{@{}l@{}}IEEE 5, 24, and \\  300 bus\end{tabular}   \\ \hline
Chen et al.~\cite{chen2017secure}        & \begin{tabular}[c]{@{}l@{}}Tsinghua University, \\ China\end{tabular}                              & 2019                   &                                                                             & \begin{tabular}[c]{@{}l@{}}Disrupt Automatic \\ Voltage Control\end{tabular}                     & Reinforcement                   & Q-learning                                  & \begin{tabular}[c]{@{}l@{}} Voltage sag- 0.5 pu \end{tabular}                     & IEEE 39 bus                                                           \\ \cline{1-3} \cline{5-9} 
Ahmadian et al.~\cite{ahmadian2018cyber}    & \begin{tabular}[c]{@{}l@{}}University of \\ Houston, USA\end{tabular}                              &                        &                                                                             & Maximize Cost                                                                                    & Unsupervised                  & GAN                                         & \begin{tabular}[c]{@{}l@{}}Generated fake \\ load data\end{tabular}                           & IEEE 5 bus                                                            \\ \cline{1-2} \cline{5-9} 
Nawaz et al.~\cite{nawaz2018machine}       & \begin{tabular}[c]{@{}l@{}}Air University, \\ Pakistan\end{tabular}                                & \multirow{-2}{*}{2018} & \multirow{-3}{*}{FDIA}                                                      & State Estimation                                                                                 & Supervised                      & LR                                          & \begin{tabular}[c]{@{}l@{}}Generated stealthy \\ FDI attack vectors\end{tabular}              & IEEE 5 bus                                                            \\ \hline
\end{tabular}
\vspace{-5pt}
\end{table*}

Until now, a few ML-based security surveys have been conducted in the smart grid domain. 
A detailed overview of ML-based security analysis of false data injection (FDI) attack in SG has been presented by Muslehet et al.~\cite{musleh2019survey}. However, the review focus was limited to a single attack. Hossain et al. conducted a study on the application of big data and ML in the electrical power grid~\cite{hossain2019application}. Most of the existing review papers do not include recent trends toward the application of ML in the security study of SG. This survey paper provides a review of state-of-the-art applications of ML in attack generation, detection, and mitigation schemes in the SG. After introducing the existing and emerging ML-based security issues, this paper attempts to inspire the researchers in providing security solutions with a view to increasing the resiliency of the SG.



\section{Machine Learning Based Attack Generation}
\label{Sec:Machine_Learning_Based_Attack_Generation}
ML-based attacks in the SG domain are less explored. Table~\ref{tab:attack} summarizes the ML-based attack generation in SG. According to the existing research works, four types of ML algorithms are utilized to generate malicious data to launch an attack in the SG. K-means and Q-learning algorithms are used to launch the line switching attacks. In contrast, Q-learning, generative adversarial networks (GAN), and linear regression (LR) models are used to generate false data injection (FDI) attacks. 

Paul et al. used load ranking and K-means clustering algorithms as two different approaches to attack SG for selecting the most vulnerable transmission lines to create contingencies~\cite{paul2019comparative}. They found that clustering-based algorithms perform better in tripping transmission lines. On the other hand, load ranking shows better results to gain higher generation loss. In~\cite{yan2016q}, Yan et al. proposed Q-learning-based cyberattacks in different buses of the system that leads the system to blackout. Ni et al. proposed another reinforcement learning-based sequential line switching attack to initiate blackout~\cite{ni2017reinforcement}. They recently proposed a multistage game using a Q-learning algorithm to create transmission line outage and generation loss~\cite{ni2019multistage}. Nawaz et al. proposed an LR-based FDI attack generator against the state estimation of the SG. They implemented and evaluated their model on IEEE 5 bus system~\cite{nawaz2018machine}. Ahmadian et al. presented a GAN-based false load data generator in~\cite{ahmadian2018cyber}. The attacker's goal was to maximize the generation cost by injecting that false load data into the system. Recently, Chen et al. also presented a Q-learning-based FDI attack generator against the automatic voltage control using partially observable Markov decision process and was able to create a voltage sag on IEEE 39 bus system~\cite{chen2018evaluation}.

\section{Machine Learning Based Attack Detection}
\label{Sec:ML_Attack_Detection}
A wide range of research has been conducted to detect various attacks in SG leveraging ML approaches. In this section, we review the existing research efforts of ML-based attack detection in various segments of the SG, as summarized in Table~\ref{tab:detection}. In the following subsections, we discuss the detection techniques of a few prevalent cyberattacks.

\begin{table*}[!ht]
\scriptsize
\centering
\caption{Classification of ML-based attack detection techniques in smart grid}
\label{tab:detection}
\begin{tabular}{|c|c|c|c|c|c|c|c|}
\hline
\textbf{Category}              & \textbf{\begin{tabular}[c]{@{}c@{}}ML \\ Algorithm\end{tabular}}                   & \textbf{Attack}                                                                      & \textbf{\begin{tabular}[c]{@{}c@{}}Number of \\Features\end{tabular}} & \textbf{Data Collection}                                                                                 & \textbf{Testbed}                                                                & \textbf{Performance}                                                                                 & \textbf{Reference}                               \\ \hline
                               &                                                                                    &                                                                                      & 304                                                                    & \begin{tabular}[c]{@{}c@{}}MATPOWER \\ simulation tool\end{tabular}                                      & IEEE 118 bus                                                                    & 99\% accuracy                                                                                        & \cite{esmalifalak2014detecting} \\ \cline{4-8} 
                               &                                                                                    &                                                                                      & NA                                                                     & NA                                                                                                       & IEEE 30 bus                                                                     & 96.1\% accuracy                                                                                      & \cite{yan2016detection}         \\ \cline{4-8} 
                               &                                                                                    & \multirow{-3}{*}{FDI}                                                               & 34                                                                     & \begin{tabular}[c]{@{}c@{}}MATPOWER \\ simulation tool\end{tabular}                                      & IEEE 14 bus                                                                     & 90.79\% accuracy                                                                                     & \cite{sakhnini2019smart}        \\ \cline{3-8} 
                               &                                                                                    & IL                                                                         & NA                                                                     & \begin{tabular}[c]{@{}c@{}}Ground truth \\ profile database\end{tabular}                                 & \begin{tabular}[c]{@{}c@{}}IEC 61850 \\ conforming testbed\end{tabular}         & 91\% accuracy                                                                                        & \cite{kaygusuz2018detection}    \\ \cline{3-8} 
                               &                                                                                    & CC                                                                                & 233                                               & SE-MF datasets                                                                                           & \begin{tabular}[c]{@{}c@{}}IEEE 14, 39-, 57 \\ and 118-bus systems\end{tabular} & \begin{tabular}[c]{@{}c@{}}99.954\% accuracy \\ and 0.939 F1-score\end{tabular}                      & \cite{ahmed2018feature}         \\ \cline{3-8} 
                               & \multirow{-6}{*}{SVM}                                                              & \begin{tabular}[c]{@{}c@{}}DoS, R2L, \\ and U2R\end{tabular}                     & NA                                                                     & NSL-KDD dataset                                                                                          & NA                                                                              & \begin{tabular}[c]{@{}c@{}}0.67\% FPR and\\ 2.15\% FNR\end{tabular}                                  & \cite{zhang2011distributed}     \\ \cline{2-8} 
                               &                                                                                    &                                                                                      & NA                                                                     & NA                                                                                                       & IEEE 30 bus                                                                     & 95.1\% accuracy                                                                                      & \cite{yan2016detection}         \\ \cline{4-8} 
                               &                                                                                    & \multirow{-2}{*}{FDI}                                                               & 34                                                                     & \begin{tabular}[c]{@{}c@{}}MATPOWER \\ simulation tool\end{tabular}                                      & IEEE 14 bus                                                                     & 85.59\% accuracy                                                                                     & \cite{sakhnini2019smart}        \\ \cline{3-8} 
                               & \multirow{-3}{*}{KNN}                                                              & CC                                                                                  & 233                                               & SE-MF datasets                                                                                           & \begin{tabular}[c]{@{}c@{}}IEEE 14, 39-, 57 \\ and 118-bus systems\end{tabular} & 77.234\% accuracy                                                                 & \cite{ahmed2018feature}         \\ \cline{2-8} 
                               & ENN                                                                                & FDI                                                                                 & NA                                                                     & NA                                                                                                       & IEEE 30 bus                                                                     & 100\% accuracy                                                                                       & \cite{yan2016detection}         \\ \cline{2-8} 
                               &                                                                                    & \begin{tabular}[c]{@{}c@{}}BF, DoS/DDoS, \\ PS, etc.\end{tabular}         & 80                                                                     & \begin{tabular}[c]{@{}c@{}}CIC-IDS2017 \\ dataset\end{tabular}                                            & NA                                                                              & 99.6\% accuracy                                                                                      & \cite{radoglou2018anomaly}      \\ \cline{3-8} 
                               & \multirow{-2}{*}{DT}                                                               & DoS                                                                     ,   & 2                                                                      & N/A                                                                                                      & N/A                                                                             & 100\%                                                                                                & \cite{hasan2019supervised}      \\ \cline{2-8} 
                               &                                                                                    & DoS                                                                                  & NA                                                                     & NS2 simulation tool                                                                                      & NA                                                                              & 99\% Accuracy                                                                                        & \cite{boumkheld2016intrusion}   \\ \cline{3-8} 
                               & \multirow{-2}{*}{NB}                                                              & CC                                                                                 & 233                                               & SE-MF datasets                                                                                           & \begin{tabular}[c]{@{}c@{}}IEEE 14, 39-, 57 \\ and 118-bus systems\end{tabular} & \begin{tabular}[c]{@{}c@{}}67.321\% accuracy \\ and 0.631 F1-score\end{tabular}                      & \cite{ahmed2018feature}         \\ \cline{2-8} 
                               & CDBfN                                                                               & FDI                                                                                 & NA                                                                     & \begin{tabular}[c]{@{}c@{}}MATPOWER \\ simulation tool\end{tabular}                                      & IEEE 118, 300 bus                                                               & 98\% accuracy                                                                                        & \cite{he2017real}               \\ \cline{2-8} 
                               &                                                                                    &                                                                                      & 48                                                                     & \begin{tabular}[c]{@{}c@{}}Irish Social Science \\ Data Archive Center\end{tabular} & NA                                                                              & 84.37\% accuracy                                                                                     & \cite{ford2014smart}            \\ \cline{4-8} 
                               &                                                                                    & \multirow{-2}{*}{FDI}                                                               & 34                                                                     & \begin{tabular}[c]{@{}c@{}}MATPOWER \\ simulation tool\end{tabular}                                      & IEEE 14 bus                                                                     & 81.78\% accuracy                                                                                     & \cite{sakhnini2019smart}        \\ \cline{3-8} 
                               & \multirow{-3}{*}{ANN}                                                              & CC                                                                                  & 233                                               & SE-MF datasets                                                                                           & \begin{tabular}[c]{@{}c@{}}IEEE 14, 39, 57 \\ and 118-bus systems\end{tabular} & \begin{tabular}[c]{@{}c@{}}86.469\% accuracy \\ and 0.863 F1-score\end{tabular}                      & \cite{ahmed2018feature}         \\ \cline{2-8} 
                               & DBsN                                                                                & FDI                                                                                 & NA                                                                     & NA                                                                                                       & \begin{tabular}[c]{@{}c@{}}IEEE 39, 118, \\ and 2848 bus\end{tabular}           & 99\% accuracy                                                                                        & \cite{karimipour2019deep}       \\ \cline{2-8} 
                               & \begin{tabular}[c]{@{}c@{}}DL model \\ (Novel)\end{tabular}                        & \begin{tabular}[c]{@{}c@{}}DoS/DDoS, \\  PS etc.\end{tabular}                     & 80                                                                     & CIC-IDS 2017 dataset                                                                                      & N/A                                                                             & 99.99\% accuracy                                                                                     & \cite{vijayanand2019novel}      \\ \cline{2-8} 
                               & EDAD                                                                               & CC                                                                                  & 233                                               & SE-MF datasets                                                                                           & \begin{tabular}[c]{@{}c@{}}IEEE 14, 39-, 57 \\ and 118-bus systems\end{tabular} & 90\% accuracy                                                                                        & \cite{ahmed2018covert}          \\ \cline{2-8} 
                               & XGBoost                                                                            & \begin{tabular}[c]{@{}c@{}}XSS, SQLI, DoS/\\ DDoS, PS, etc.\end{tabular} & 71                                                                     & CIC-IDS2018 dataset                                                                                      & NA                                                                              & \begin{tabular}[c]{@{}c@{}}99.87\% precision \\ and  99.75\% recall\end{tabular}                     & \cite{roy2019network}           \\ \cline{2-8} 
                               & \begin{tabular}[c]{@{}c@{}}DT coupled \\ SVM (Novel)\end{tabular}                  & ET                                                                                   & NA                                                                     & NA                                                                                                       & NA                                                                              & 92.5\% accuracy                                                                                      & \cite{jindal2016decision}       \\ \cline{2-8} 
                               & Adaboost                                                                           & CC                                                                                 & 233                                               & SE-MF datasets                                                                                           & \begin{tabular}[c]{@{}c@{}}IEEE 14, 39-, 57 \\ and 118-bus systems\end{tabular} & \begin{tabular}[c]{@{}c@{}}85.958\% accuracy \\ and 0.852 F1-score\end{tabular}                      & \cite{ahmed2018feature}         \\ \cline{2-8} 
                               & AIRS                                                                      & \begin{tabular}[c]{@{}c@{}}DoS, R2L, \\ U2R, and PA\end{tabular}                     & NA                                                                     & NSL-KDD dataset                                                                                          & NA                                                                              & \begin{tabular}[c]{@{}c@{}}1.3\% FPR, and\\ 26.32\% FNR\end{tabular}                                 & \cite{zhang2011distributed}     \\ \cline{2-8} 
                               & Multi-SVM                                                                          & \begin{tabular}[c]{@{}c@{}}DoS \end{tabular}          & 44                                                                     & ADFA-LD                                                                                                  &  NA                                                                               & 90 accuracy\%                                                                                                 & \cite{prasad2019machine}        \\ \cline{2-8}
                               
                               & Autoencoder ANN                                                                          & \begin{tabular}[c]{@{}c@{}}FDI.\end{tabular}          & 339                                                                    & NA                                                                                                 &  IEEE 118 bus                                                                               & 95.05\% accuracy                                                                                                 & \cite{wang2020detection}        \\
                               \cline{2-8}
                               
                               & NARX ANN                          & \begin{tabular}[c]{@{}c@{}}FDI.\end{tabular}          & NA                                                                    & OPAL-RT simulator                                                                                                 &  DC microgrid system                                                                               & 95.05\% accuracy                                                                                                 & \cite{habibi2020detection}        \\
                               \cline{2-8}
                               &                                                                                    &                                                                                      & 112                                                                    & \begin{tabular}[c]{@{}c@{}}MATPOWER \\ simulation tool\end{tabular}                                      & IEEE 30 bus                                                                     & \begin{tabular}[c]{@{}c@{}}99.9\% accuracy,\\ 91.529\% precision, \\ and 85.02\% recall\end{tabular} & \cite{ayad2018detection}        \\ \cline{4-8} 
\multirow{-28}{*}{Supervised}  & \multirow{-2}{*}{RNN}                                                              & \multirow{-2}{*}{FDI}                                                               & 41                                                                     & NSL-KDD dataset                                                                                          & IEEE 39 bus                                                                     & 90\% accuracy                                                                                        & \cite{niu2019dynamic}           \\ \hline
                               & Statistical                                                                        & FDI                                                                                 & 304                                                                    & \begin{tabular}[c]{@{}c@{}}MATPOWER \\ simulation tool\end{tabular}                                      & IEEE 118 bus                                                                    & 99\% accuracy                                                                                        & \cite{esmalifalak2014detecting} \\ \cline{2-8} 
                               & \begin{tabular}[c]{@{}c@{}}ART and SOM \\ based classifier \\ (Novel)\end{tabular} & FDI                                                                                 & 24                                                                     & 
                               \begin{tabular}[c]{@{}c@{}}Real-time digital  \\Simulator(RTDS)\end{tabular}                                                                                            & \begin{tabular}[c]{@{}c@{}}RTDS hardware \\ in the loop testbed\end{tabular}    & 90\% accuracy                                                                                        & \cite{valdes2016anomaly}        \\ \cline{2-8} 
                               & CLONALG                                                                            & \begin{tabular}[c]{@{}c@{}}DoS, R2L, \\ U2R\end{tabular}                     & NA                                                                     & NSL-KDD dataset                                                                                          & NA                                                                              & \begin{tabular}[c]{@{}c@{}}0.7\% FPR, and\\ 21.02\% FNR\end{tabular}                                 & \cite{zhang2011distributed}     \\ \cline{2-8} 
\multirow{-4}{*}{Unsupervised} & iForest                                                                            & CC                                                                               & 233                                               & SE-MF datasets                                                                      & \begin{tabular}[c]{@{}c@{}}IEEE 14, 39-, 57 \\ and 118-bus systems\end{tabular} & 90\% accuracy                                                                                        & \cite{ahmed2019unsupervised}    \\ \hline
\end{tabular}
\vspace{-5pt}
\end{table*}

\begin{table*}[!htb]
\centering
\scriptsize
\caption{Classification of ML-based attack mitigation techniques in smart grid\label{tab:mitigation}}
\begin{tabular}{|l|l|r|l|l|l|l|}
\hline
\textbf{Reference}                                         & \textbf{Institution}                                                                    & \multicolumn{1}{l|}{\textbf{\begin{tabular}[c]{@{}l@{}}Publication \\ Year\end{tabular}}} & \textbf{Attack}       & \textbf{\begin{tabular}[c]{@{}l@{}}ML Model \\ Type\end{tabular}} & \textbf{\begin{tabular}[c]{@{}l@{}}ML \\ Algorithm\end{tabular}} & \textbf{Testbed}                                                          \\ \hline
Chen et al. \cite{chen2018evaluation}     & \begin{tabular}[c]{@{}l@{}}Tsinghua University Beijing,\\  China\end{tabular}           & 2019                                                                                      & \multirow{3}{*}{FDI} & Reinforcement                                                     & Q-learning                                                       & IEEE 39 bus                                                               \\ \cline{1-3} \cline{5-7} 
Li et al. \cite{li2019online}             & \begin{tabular}[c]{@{}l@{}}North China Electric Power \\ University, China\end{tabular} & 2019                                                                                      &                       & \multirow{2}{*}{Unsupervised}                                     & GAN                                                              & \begin{tabular}[c]{@{}l@{}}IEEE 30, and \\ 118 bus\end{tabular}           \\ \cline{1-3} \cline{6-7} 
Wei et al. \cite{wei2016deep}             & University of Akron, USA                                                                & 2016                                                                                      &                       &                                                                   & DBfN                                                             & \begin{tabular}[c]{@{}l@{}}New England 39\\ bus power system\end{tabular} \\ \hline
An et al. \cite{an2019defending}            & \begin{tabular}[c]{@{}l@{}}Xian Jiaotong University, \\ China)\end{tabular}             & 2019                                                                                      & \multirow{2}{*}{DIA}  & Reinforcement                                                     & Q-learning                                                       & \begin{tabular}[c]{@{}l@{}}IEEE 9, 14, and \\ 30 bus system\end{tabular}  \\ \cline{1-3} \cline{5-7} 
Parvez et al. \cite{parvez2016securing}   & \begin{tabular}[c]{@{}l@{}}Florida International University,\\ USA\end{tabular}         & 2016                                                                                      &                       & Supervised              & KNN                                                                                  & AMI network                                                               \\ \cline{1-4} \cline{5-7} 
Mahrjan et al. \cite{maharjan2019machine} & \begin{tabular}[c]{@{}l@{}}University of Texas  at Dallas,\\ USA\end{tabular}           & 2019                                                                                      & \multirow{2}{*}{DUA}  &                           Unsupervised                                        & SVM                                                              & MPEI                                                                      \\ \cline{1-3} \cline{6-7} 
Ren et al. \cite{ren2019fully}            & \begin{tabular}[c]{@{}l@{}}Nanyang Technological \\ University, Singapore\end{tabular}  & 2019                                                                                      &                       &                                                                   & GAN                                                              & \begin{tabular}[c]{@{}l@{}}New England 39 \\ bus\end{tabular}             \\ \cline{1-4} \cline{6-7} 
Shahriar et al. \cite{shahriar2020g}      & Florida International University                                                        & 2020                                                                                      & \multirow{2}{*}{LAD}  &                                                                   & GAN                                                              & KDD-99                                                                    \\ \cline{1-3} \cline{6-7} 
Ying et al. \cite{ying2019power}          & Zhejiang University, China                                                              & 2019                                                                                      &                       &                                                                   & GAN                                                              & Synthetic                                                                 \\ \hline
\end{tabular}
\vspace{-5pt}
\end{table*}

\subsection{False Data Injection Attack}
Most of the research efforts attempted to detect stealthy FDI attacks using ML. Esmalifalak et al. attempted to detect stealthy FDI attacks using a support vector machine (SVM)-based technique and a statistical anomaly detection approach~\cite{esmalifalak2014detecting}. They showed that SVM outperforms the statistical approach when the model is trained with sufficient data. He et al. proposed a conditional deep belief network (CDBfN)-based detection scheme that extracts temporal features from distributed sensor measurements~\cite{he2017real}. The proposed detection scheme is robust against various attacked measurements and environmental noise levels. Moreover, it can perform better than SVM and artificial neural network (ANN)-based detection mechanisms. Karimipour et al. proposed a continuous, computationally efficient, and independent mechanism using feature extraction scheme and time series partitioning method to detect FDI attacks~\cite{karimipour2019deep}. This paper used dynamic bayesian networks (DBsN) concept and Boltzmann machine-based learning algorithms to detect unobservable attacks. Valdes et al. presented a novel intrusion detection system (IDS) utilizing adaptive resonance theory (ART) and self-organizing maps (SOM) to differentiate normal, fault, and attack states in a distribution substation system~\cite{valdes2016anomaly}. Yan et al. viewed the FDI attack detection problem as a binary classification problem and attempted to solve it using three different algorithms: SVM, K-nearest neighbor (KNN), and extended nearest neighbor (ENN)~\cite{yan2016detection}. Their experimental analysis showed that all these algorithms could be tuned to attain optimal performance against FDI attack detection. Ayad et al. examined the use of a recurrent neural network (RNN)-based method that deals with temporal and spatial correlations between the measurements, unlike other learning methods, to recognize FDI attacks in SG.~\cite{ayad2018detection}. Niu et al. presented a deep learning-based framework combining a convolutional neural network (CNN) and a long short term memory (LSTM) network to detect novel FDI attacks~\cite{niu2019dynamic}. Sakhnini et al. analyzed three different algorithms (e.g., ANN, KNN, and SVM) that incorporate different feature selection (FS) techniques and used a genetic algorithm (GA) as optimal FS method for power systems. The authors of~\cite{habibi2020detection} proposed a nonlinear autoregressive exogenous (NARX) neural networks to estimate DC current and voltage to detect FDI attacks in DC microgrid and showed that the proposed method has successfully identified FDI attacks during transient and steady-state conditions. The autoencoder ANN-based FDI attack detection mechanism has also been investigated 
by Wang et al.~\cite{wang2020detection}.
\subsection{Covert Cyber Attack}
Ahmed et al. tried to detect covert cyber (CC) attacks in their several research efforts. In one of their works, they proposed two euclidean distance-based abnormality recognition scheme for detecting anomalies in the state estimation measurement features (SE-MF) dataset~\cite{ahmed2018covert}. In their another work, they leveraged several ML methods (KNN, SVM, multi-layer perceptron (MLP), naïve bayes (NB), and adaboost) to identify a CC attack in the SE information that is gathered through a communication network of smart grid~\cite{ahmed2018feature} along with a GA for optimizing the features. Their discovery revealed that KNN has low CC attack detection performance than the other ML methods. They also proposed an unsupervised ML-based mechanism utilizing a state-of-the-art algorithm, called isolation forest (iForest) to distinguish CC attacks in SG systems using non-labeled information. The proposed mechanism can sensibly improve detection performance in the periodic operational condition~\cite{ahmed2019unsupervised}.


\subsection{Electricity Theft}
Energy Theft (ET) detection in SG mostly leverages supervised ML algorithms. Ford et al. examined a novel utilization of ANN and smart meter fine-grained information to report energy fraud, accomplishing a higher energy theft detection rate ~\cite{ford2014smart}. Jindal et al. proposed an extensive top-down scheme utilizing a decision tree (DT) and SVM. In contrast to other works, the proposed scheme is sufficiently proficient in accurately distinguishing and detecting constant power theft at each level in the power transmission system~\cite{jindal2016decision}.

\subsection{Denial of Service Attack}
Denial of Service (DoS) and Distributed Denial of Service (DDoS) attack detection has got a significant research focus. Vijayanand et al. proposed a novel DoS attack detection framework utilizing diverse multi-layer deep learning algorithms for precisely identifying the attacks by analyzing smart meter traffic~\cite{vijayanand2019novel}. In another work, they have used another novel approach named as multi-SVM for DoS attack detection~\cite{prasad2019machine}. Zhang et al. attempted to detect anomalies in diverse layer network structure using SVM, clonal selection algorithm (CLONALG), and artificial immune recognition system (AIRS). 
According to their performance analysis, SVM based IDS outperforms CLONALG and AIRS~\cite{zhang2011distributed}. Radoglou et al. proposed a novel IDS for AMI using DT\cite{radoglou2018anomaly}. Boumkheld et al. proposed an NB classifier based centralized IDS for accumulating all information sent by data collector that requires massive memory and computational resources. Roy et al. used extreme gradient boosting (XGBoost), random forest (RF) on CIC-IDS2018 dataset for detecting various SG cyberattacks, including DoS~\cite{roy2019network}.



Moreover, the ML-based detection of cross-site scripting (XSS) and SQL injection (SQLI), unknown routing attack (URA), brute force (BF), information leakage (IL), port scanning (PS), remote to local (R2L), the user to root (U2R) attacks also gained some attention in the recent research. 

\section{Machine Learning Based Attack Mitigation}
\label{Sec:Machine_Learning_Based_Attack_Mitigation}

\begin{figure*}[t]
    \begin{center}
     \subfigure[]
        {
        \label{gen}
            \includegraphics[width=0.32\textwidth, keepaspectratio=true]{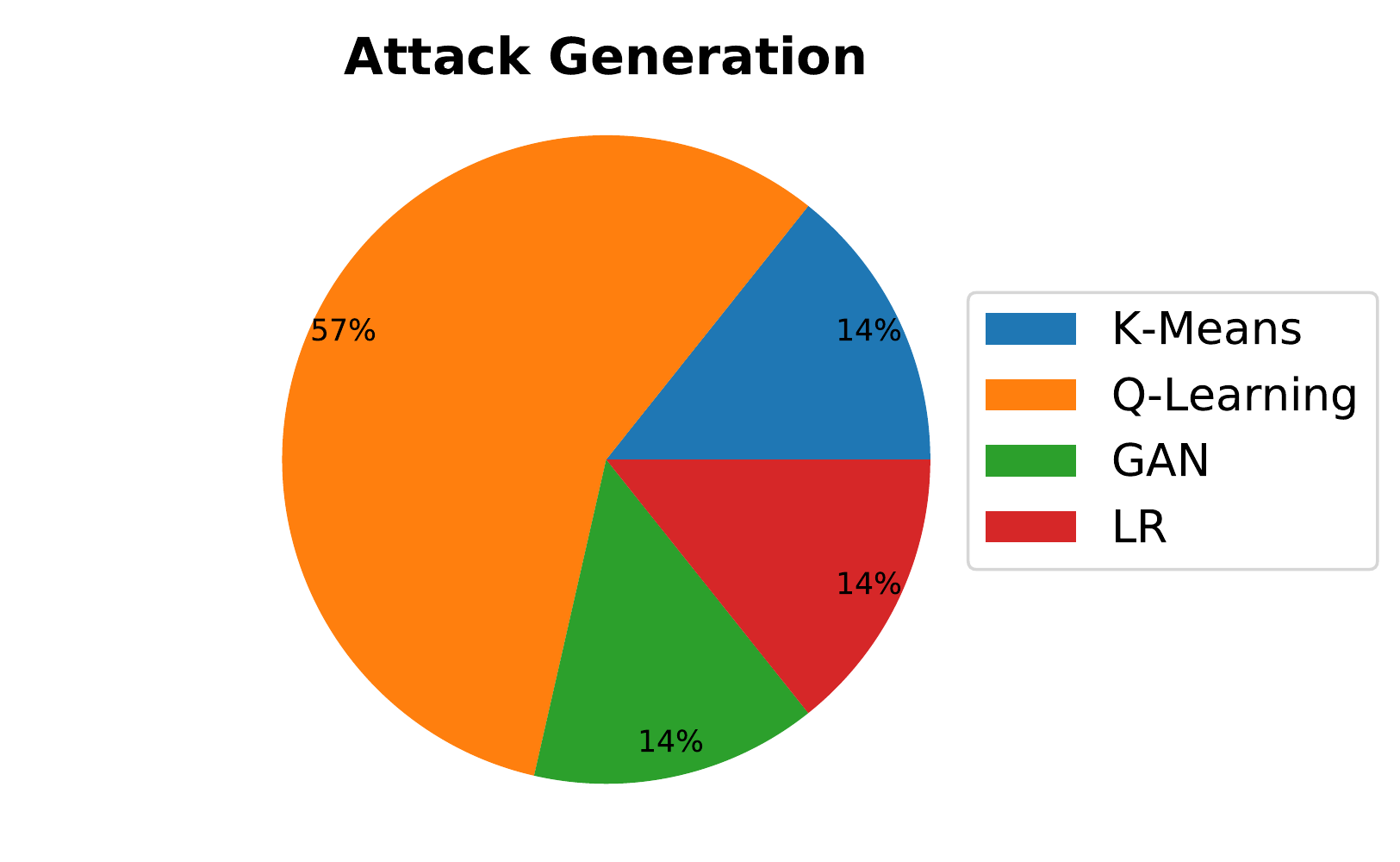}
        }
        \hspace{-15pt}
     \subfigure[]
        {
        \label{det}
           \includegraphics[width=0.32\textwidth, keepaspectratio=true]{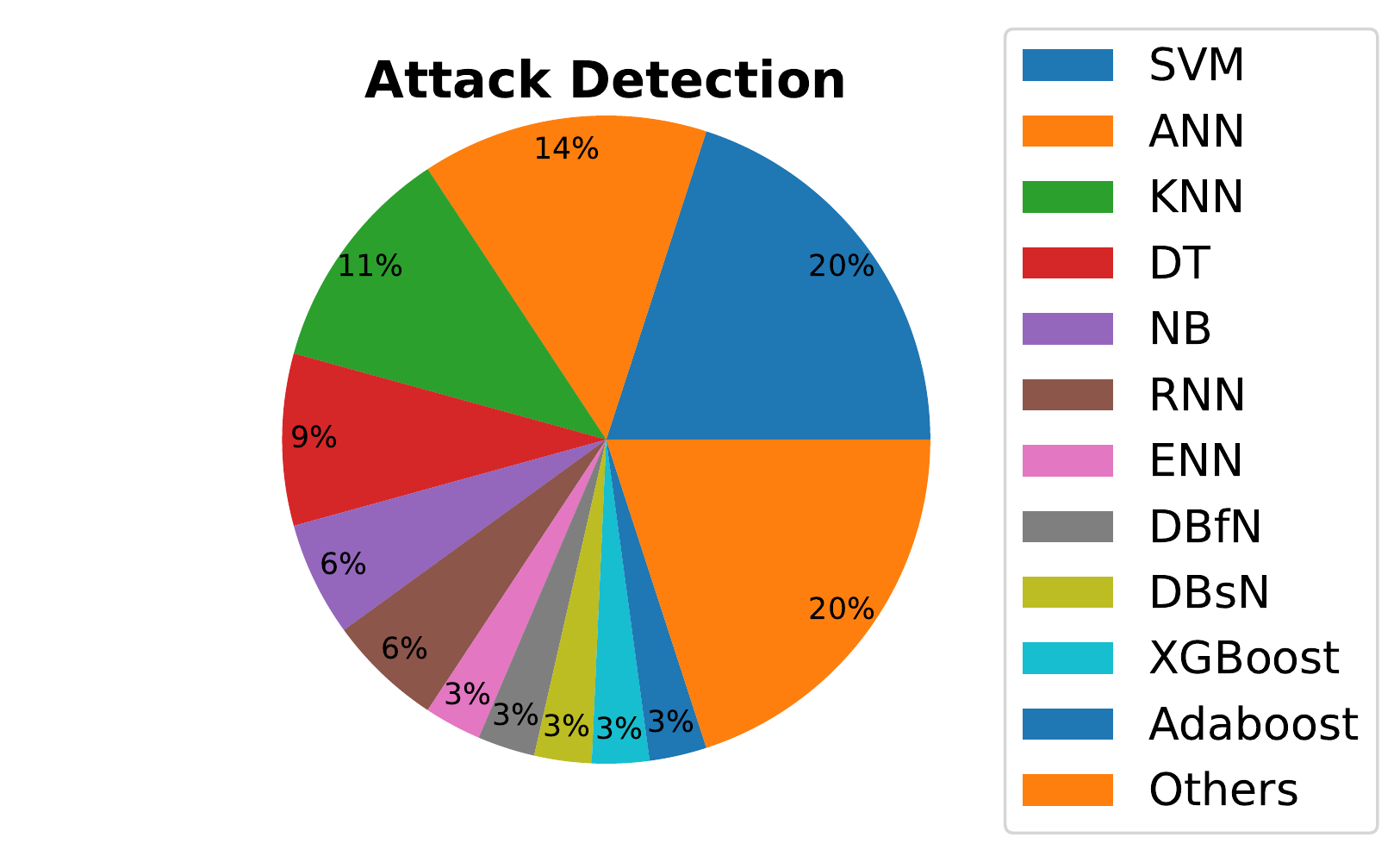}
        }
        \hspace{-15pt}
        \subfigure[]
        {
        \label{mit}
           \includegraphics[width=0.32\textwidth, keepaspectratio=true]{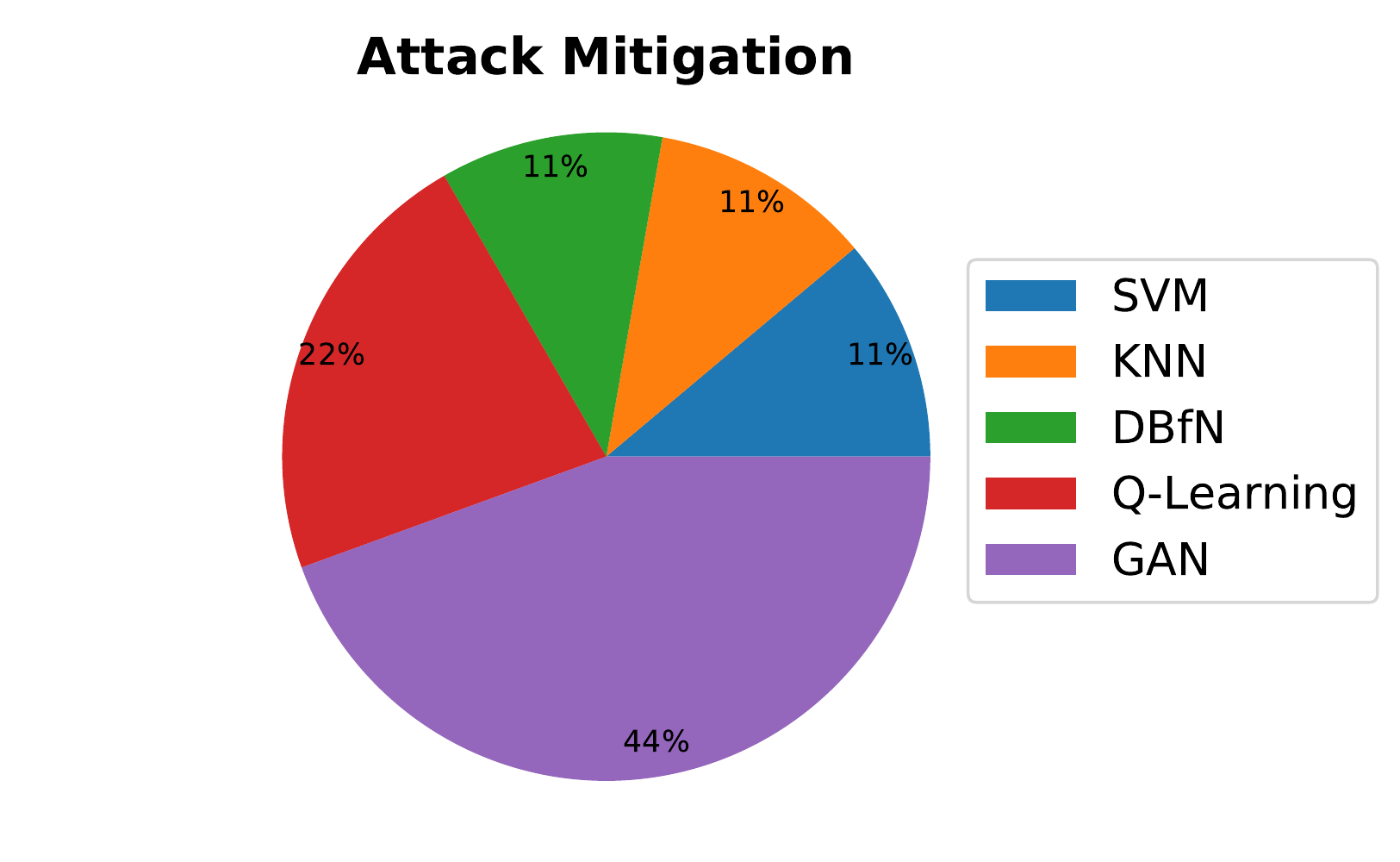}

        }              
        \hspace{-15pt}
    \end{center}
    \vspace{-10pt}
    \caption{Pie-chart of mostly used ML techniques in a) generation, b) detection, and  c) mitigation of cyberattacks in smart grid}
    \label{Fig:piechart}
\vspace{-10pt}
\end{figure*}


Attack mitigation is the strategy to minimize the effect of malicious attacks maintaining the functionality of the system.
Table~\ref{tab:mitigation} summarizes the ML-based attack mitigation in SG.

Wei et al. proposed a deep belief network (DBfN)-based cyber-physical model to identify and mitigate the FDI attack while maintaining the transient stability of wide-area monitoring systems (WAMSs)~\cite{wei2016deep}. An et al. modeled a deep-Q-network (DQN) detection scheme to defend against data integrity attacks (DIA) in AC power systems~\cite{an2019defending} and showed that the DQN detection outperforms the baseline schemes in terms of detection accuracy and rapidity. Chen et al. presented a Q-learning-based mitigation technique for FDI attacks in automatic voltage controller~\cite{chen2018evaluation}. They replaced the suspected data with their maximum likelihood estimation (MLE) values to enhance the securities of the state estimation and OPF-based controls. In~\cite{maharjan2019machine}, Maharjan et al. proposed an SVM based resilient SG network with DERs to mitigate the data unavailability attack (DUA). Parvez et al. proposed a localization-based key management system using the KNN algorithm for node/meter authentication in AMI networks~\cite{parvez2016securing}. They showed that the source meter could be authenticated accurately by the KNN algorithm utilizing the pattern of sending frequency, packet size, and distance between two meters. Ren et al. proposed a GAN model that predicts the missing/unavailable PMU data even without observability and network topology~\cite{ren2019fully}. Shahriar et al. proposed a GAN-based approach to generate a synthetic attack dataset from the existing attack data. They achieved up to 91\% f1 score in detecting different cyberattacks for the emerging smart grid technologies~\cite{shahriar2020g}. In another work, Ying et al. proposed a similar GAN-based approach achieving a 4\% increase in the attack detection accuracy~\cite{ying2019power}. Li et al. proposed another GAN based model to defend against FDI attacks~\cite{li2019online} that provides the predicted deviation of the measurements and recovers the compromised sensors.



  
    


\section{Future Research Direction}
\label{Sec:future_research_direction}
Fig.~\ref{Fig:piechart} shows the pie-chart of ML techniques used in attack generation, detection, and mitigation for the SG network. Fig.~\ref{det} illustrates that a lot of research have been conducted towards the application of different ML algorithms in attack-detection, whereas, generation and mitigation fields are comparatively less explored. As SG is dynamic and intermittent in nature, researchers are mostly applying Q-learning to generate real-time attacks, as shown in Fig.~\ref{gen}. On the other hand, GAN has the capability of generating missing data with considerable accuracy, thus, as shown in Fig.~\ref{mit}, it is prominently used in attack mitigation strategies. However, GAN also has the potential to generate attack data considering the system's topology and states. Hence, future researchers can focus on GAN and the other less explored algorithms such as ANN, RNN, KNN, DT, etc. in the attack generation and mitigation strategies.  

\section{Conclusion}
\label{Sec:Conclusion}
ML has been creating new dimensions for both attackers and defenders with respect to scalability and accuracy due to its wide range of applications in SG. 
Therefore, it draws attention to the researchers for conducting security-related investigations applying emerging ML algorithms. In this paper, we review current research works, related to the potential ML-based attack generation, detection, and mitigation schemes for future researchers. In addition, we  present a tabular form summarizing the existing studies in an organized way that would help future researchers to emphasize the unfocused areas.

\section{Acknowledgement}
\label{Sec:Acknowledgement}
This research was supported in part by the U.S. National Science Foundation under awards \#1553494 and \#1929183.

\bibliographystyle{unsrt}
\bibliography{main_paper} 
\end{document}